# Electron spin-lattice relaxation of $Er^{3+}$-ions in $Y_{0.99}Er_{0.01}Ba_2Cu_3O_x$

V.A. Ivanshin [a,b,*], M.R. Gafurov [a], I.N. Kurkin [a], S.P. Kurzin [a], A. Shengelaya [b], H. Keller [b], M. Gutmann [c]

[a] *MRS Laboratory, Kazan State University, 420008 Kazan, Russian Federation*
[b] *Physik-Institut der Universität Zürich-Irchel, CH-8057 Zürich, Switzerland*
[c] *Laboratory for Neutron Scattering, ETH Zürich and Paul Scherrer Institut, CH-5232 Villigen, Switzerland*

**Abstract**

The temperature dependence of the electron spin-lattice relaxation (SLR) was studied in $Y_{0.99}Er_{0.01}Ba_2Cu_3O_x$ ($0 \leq x \leq 7$). The data derived from the electron spin resonance (ESR) and SLR measurements were compared to those from inelastic neutron scattering studies. SLR of $Er^{3+}$-ions in the temperature range from 20 K to 65 K can be explained by the resonant phonon relaxation process with the involvement of the lowest excited crystalline-electric-field electronic states of $Er^{3+}$. These results are consistent with a local phase separation effects. Possible mechanisms of the ESR line broadening at lower temperatures are discussed. 1998 Physica C 307 p.61-66.

## 1. Introduction

The measurements of nuclear- and electron spin-lattice relaxation (SLR) time $T_1$ can provide an useful information about electronic states and internal fields in high-$T_c$ superconductors (HTSC) [1–9]. The electron relaxation of different paramagnetic centers, either impurities (such as $Gd^{3+}$, $Fe^{3+}$, $Yb^{3+}$) [2–8] or belonging to the host lattice ($Cu^{2+}$) [9] was studied in the perowskite-type compounds $YBa_2Cu_3O_x$ (YBCO) ($6 \leq x \leq 7$) by means of measuring relaxation times from electron spin resonance (ESR) linewidth [2–7] or directly using the relation between the ESR absorption magnitude and the response of longitudinal spin magnetization [8,9]. The physical properties of YBCO are very sensitive to the oxygen content $x$, determining the occurrence ($6.4 \leq x \leq 7$) and disappearance of superconductivity for $x < 6.4$. For $6 \leq x \leq 6.5$ the Cu ions align antiferromagnetically, and the interplay between magnetism and superconductivity could be observed. Several reasons caused our choice of the $Er^{3+}$ ion as a paramagnetic dopant in YBCO. Values of $g$-factors of this ion are strongly determined by the symmetry of crystalline-electric-field (CEF). ESR spectrum of $Er^{3+}$ ($S_{eff} = 1/2$; $g \neq 2$) is very simple and is situated far from an unavoidable $Cu^{2+}$-impurity's signal with $g \sim 2$. ESR spectra lines of $Er^{3+}$ ($g > 2$) have usually much stronger intensities than those with a smaller $g$, but their linewidth is small enough for the SLR measurements in a wide temperature range. Since the discovery of HTSC there are few reports, devoted to ESR of $Er^{3+}$ in YBCO [10–16]. The special attention in our investigation is given to the study of the temperature dependence of the ESR linewidth and possible mechanisms of the electron SLR. One of the aims of this work is also to compare our SLR data with results of inelastic neutron scattering (INS) measurements on similar samples [17–19], which were obtained using the same sample preparation technique.

---

* Corresponding author. Kazan State University, MRS Laboratory, Kremlevskaya str. 18, 420008 Kazan, Russia.
E-mail: vladimir.ivanshin@ksu.ru

## 2. Sample preparation

The polycrystalline $Y_{0.99}Er_{0.01}Ba_2Cu_3O_x$ samples were prepared in a conventional solid state reaction (at the Kazan University) and via sol–gel method [20] (in the Laboratory for Neutron Scattering of ETH Zürich and Paul Scherrer Institut). The sol–gel method was applied to ensure homogeneous distribution of Er on the Y site. Starting materials were the metal nitrates. These were dissolved in stoichiometric amounts in demineralized water. The citric acid and ethylenglycol were added to form a polymer with the metal ions uniformly embedded. The liquid was evaporated below 90°C under continuous stirring. The obtained gel then was dried at around 120°C to obtain the precursor. The precursor was heated at 300, 500 and 700°C in order to burn all organic remains. The powder was ground and pressed into pellets. Single phase samples with $x = 7$ were obtained after sintering at 890, 900, 910, 920, and 930°C. Several smaller batches were reduced in order to obtain all other oxygen stoichiometries. No contaminant phases were revealed by X-ray measurements.

ESR of the non-oriented powder HTSC samples is rather complicated and as a rule, is attributed with a certain indefinity of measurements because of a large difference of $g$-tensor components and related broadening of ESR lines. Therefore, the quasi-single-crystal samples were prepared, where the powders were mixed with epoxy resin or paraffin and were placed in a glass cells in a strong magnetic field until the resulting suspension hardened. The value of this constant magnetic field was varied in a range from 1.5 to 9 $T$ to become ESR lines so narrow as it was possible. As a result of this treatment, the $c$ axes of the individual crystallites were aligned along the direction of external magnetic field.

## 3. Results and discussion

The ESR spectra were recorded at a $X$ band (9.25 GHz) IRÉS-1003 spectrometer in a temperature range from 4 to 70 K. A low temperature, well resolved ESR spectrum of $Er^{3+}$ (Fig. 1a, b) can be described by a spin Hamiltonian of either axial ($x < 6.5$) or rhombic ($x > 6.5$) symmetry [11]. In the latter case we can assume, that $g_x = g_y = g_\perp$, $g_z = g_\parallel$, because the difference between $g$-tensor components in the $xy$ plane is small. No remarkable changes of the ESR linewidth were observed near $T_c$. The values of $g$-factors measured at $T = 30$ K (Table 1) depend weakly on the oxygen content $x$ and temperature. The $g$-factors, which were calculated due to simulations of ESR transition of $Er^{3+}$ in YBCO using the CEF coefficients, determined from inelastic neutron scattering [18], are listed in Table 2. They coincide very well with the values from ESR measurements.

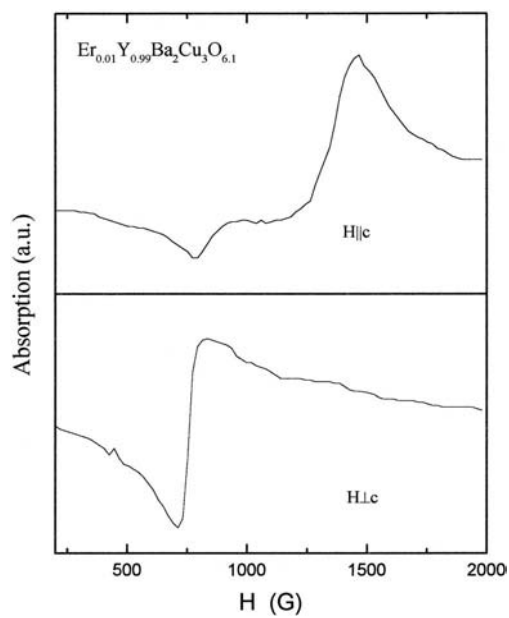

Fig. 1. $Er^{3+}$-ESR spectrum in $Y_{0.99}Er_{0.01}Ba_2Cu_3O_{6.1}$ at $T = 4.2$ K for both orientations of external magnetic field $H$; at the top, $H \parallel c$ and at the bottom, $H \perp c$.

Table 1
Effective $g$-values, residual ESR linewidth $\Delta H_{pp}^{RES}$ for both orientations of external magnetic field extracted for the various samples and the SLR parameters $B_{1,2}$ and $\Delta_{1,2}$ determined by the fitting of relation (3) from the temperature dependence of the ESR linewidth

| $x$ | 6.85 | 6.59 | 6.46 | 6.29 | 6.12 | 6.1 | 6.0 |
|---|---|---|---|---|---|---|---|
| $T_c$ (K) | 85 | 58 | 48 | – | – | – | – |
| $g_\parallel$ | 4.3(1) | 4.6 | ∼ 4.5 | 4.95 | 4.9(1) | 4.76 | 4.9(1) |
| $g_\perp$ | 7.6(1) | 7.3 | 7.13(10) | 7.2 | 7.15(15) | 7.2 | 7.15(15) |
| $\Delta H_{pp\parallel}^{res}$ (G) | 230 | 440 | ≥ 1000 | 370 | 210 | 290 | 190 |
| $\Delta H_{pp\perp}^{res}$ (G) | 120 | 120 | 110 | 150 | 175 | 190 | 170 |
| $B_{1,2}$ (s$^{-1}$) | $1 \times 10^{11}$ | – | $1.3 \times 10^{11}$ | – | $(3.9; 15.2) \times 10^{10}$ | – | $(3.9; 15.2) \times 10^{10}$ |
| $\Delta_{1,2}/k_B$ (K) | 108(11) | – | 125(12) | – | 80(8); 120(10) | – | 80(8); 120(10) |

We have observed a rapid increase of the ESR linewidth, faster than linear, with increasing temperature above 15 K. ESR spectrum disappears at $T \geq 65$ K. The similar behaviour of the $Er^{3+}$ ESR linewidth, which is caused by the 4f-electron–phonon interaction [21], was observed both in YBCO [7,13,14,16] and $La_{2-x}Sr_xCuO_{4-\delta}$ [15,22]. However, in contrast to results reported recently by Shimizu et al. about the Lorentzian line shape on their non-oriented powdered YBCO samples [14,16], all our ESR experiments have revealed rather a complicated superposition of Lorentzian and Gaussian line shapes. As predicts in this case the line shape analysis of complicated ESR spectra [23], the ESR linewidth, caused only by the SLR and interpreted usually as a Lorentzian line, $\Delta H_{pp}^{SLR}$, can be extracted from the experimentally measured ESR linewidth $\Delta H_{pp}$ using the relation:

$$(\Delta H_{pp})^2 = \Delta H_{pp}^{SLR} \Delta H_{pp} + (\Delta H_{pp}^{RES})^2, \quad (1)$$

where $\Delta H_{pp}^{RES}$ is the minimal (residual) ESR linewidth. Preliminary ESR studies of the residual linewidth at $X$- and $Q$-band on the sample with $x = 6.85$ [7,11] have revealed the proportional ESR line broadening with frequency increasing, which is attributed usually to a Gaussian lineshape. The SLR rate $T_1^{-1}$ is defined by expression [24]:

$$T_1^{-1}(s^{-1}) = (\sqrt{3}/2) g \mu_B \hbar \Delta H_{pp}^{SLR} (G)$$
$$\equiv 7.62 \times 10^6 g \Delta H_{pp}^{SLR}, \quad (2)$$

where $\mu_B$ is Bohr's magneton and $\hbar$ is a Planck's constant. An equality of electron SLR $T_1^{-1}$ and spin–spin $T_2^{-1}$ relaxation rates is assuming in this approach.

The temperature dependence of the $Er^{3+}$-linewidth, which is presented in Fig. 2 for $x = 6.46$, has two main contributions, $\Delta H_{pp}^{RES}$ and $\Delta H_{pp}^{SLR}$. The values of $\Delta H_{pp}^{RES}$, determined in the low temperature range from 4.2 K to 15 K for both orientations of the external magnetic field, are listed in Table 1 and do not show any kind of monotonous dependence on $x$. However, only additional ESR studies at different microwave frequencies can separate the inhomogeneous part of the ESR line broadening caused by the difference of the CEF parameters from the homogeneous one in order to investigate the charge distribution of the $Er^{3+}$-ions between copper-oxide planes.

$\Delta H_{pp}^{SLR}$ depend exponentially on temperature and the corresponding SLR rate $T_{1O}^{-1}$ can be written as:

$$T_{1O}^{-1} = B_1 \exp(-\Delta_1/T) + B_2 \exp(-\Delta_2/T), \quad (3)$$

where parameters $B_{1,2}$ and $\Delta_{1,2}$ are also given in Table 1. The dependence of this kind corresponds to the Orbach–Aminov process (resonant phonon relaxation process) [25], and here $\Delta_{1,2}$ are the energy separation between the ground and excited electronic states, and parameters $B_{1,2}$ depend normally on the strength of the orbit–lattice coupling. The temperature dependence of the SLR rate $T_1^{-1}$ for three different oxygen contents is shown in Fig. 3. The significant ESR line broadening prevented our SLR measurements for $x = 6.29$ and $6.59$. We found no dependence of the SLR rate on orientation, as it is presented in Fig. 4 for $x = 6.12$.

Table 2
Effective $g$-values calculated from CEF energy scheme of term $^4I_{15/2}$ of $Er^{3+}$ in $ErBa_2Cu_3O_x$ and the most intensive CEF excitations ($A_i$, $C$) determined from INS measurements [18] against oxygen content $x$

| $x$ | 6.98 | 6.45 | 6.09 | 6.0 |
|---|---|---|---|---|
| $g_\parallel$ | 4.28 | 4.64 | 4.89 | 4.89 |
| $g_\perp$ | 7.69 | 7.57 | 7.45 | 7.39 |
| $A_i, C$ (K) | 105.8($A_1$) | 125.2($C$) | 79.4($A_3$); 125.2($C$) | – |

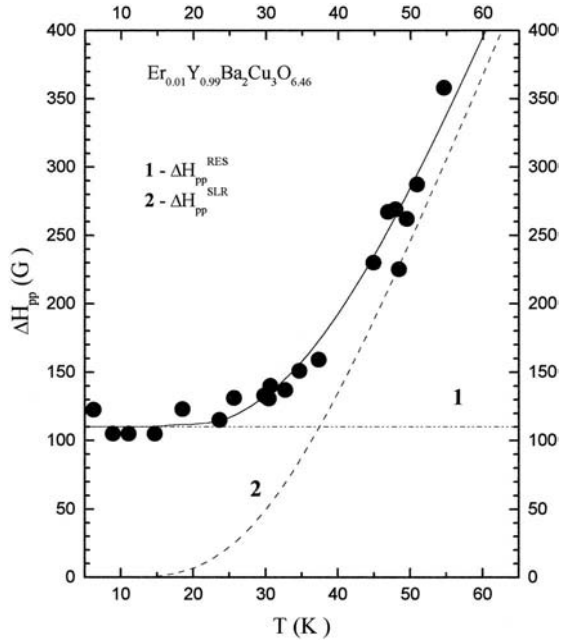

Fig. 2. Temperature dependence of the $Er^{3+}$-ESR linewidth in $Y_{0.99}Er_{0.01}Ba_2Cu_3O_{6.46}$. Dashed lines are contributions from the residual ESR linewidth (1) and the Orbach SLR (2); solid line is a sum of both contributions.

The values of $\Delta_{1,2}$, measured from SLR, depend on oxygen content and coincide quite good with either one or two lowest and most intensive CEF excitations ($A_i$, $C$) determined from INS measurements (Table 2), in which the full CEF energy scheme of $Er^{3+}$ ion in the ground-state J multiplet in $ErBa_2Cu_3O_x$ was studied [18]. The energy states $A_i$ are associated with different clusters (local regions) of metallic ($A_1 \simeq 105.8$ K; $A_2 \simeq 99.6$ K) or semiconducting ($A_3 \simeq 79.4$ K) character [19]. The intensity of corresponding transitions depends strongly on oxygen content and arrives maximum weight close to $x = 7.0$ (for $A_1$), 6.5 ($A_2$) and 6.0 ($A_3$), respectively. It is remarkable that only the most intensive CEF energy excitation $C \simeq 125.2$ K is involved in the SLR processes in the superconducting sample with $x = 6.46$, where intensities of all lower $A_i$ transitions are extremely weak. These results are significantly different from those presented by Shimizu et al., in which the temperature dependence of the $Er^{3+}$-ESR linewidth in YBCO was investigated and the monotonous decreasing of $\Delta$ from 140 K at 6.2 to approximately 110 K close to oxygen concentration of 7, was reported [14,16]. We believe that our SLR measurements provide an additional experimental evidence that observed CEF excitations originate from different local environments of the $Er^{3+}$-ions, indicating local phase separation in $Y_{0.99}Er_{0.01}Ba_2Cu_3O_x$.

Theory of the Orbach SLR predicts for the relaxation between ground doublet and excited Kramers doublet, if the crystal field splitting $\Delta$ is less than

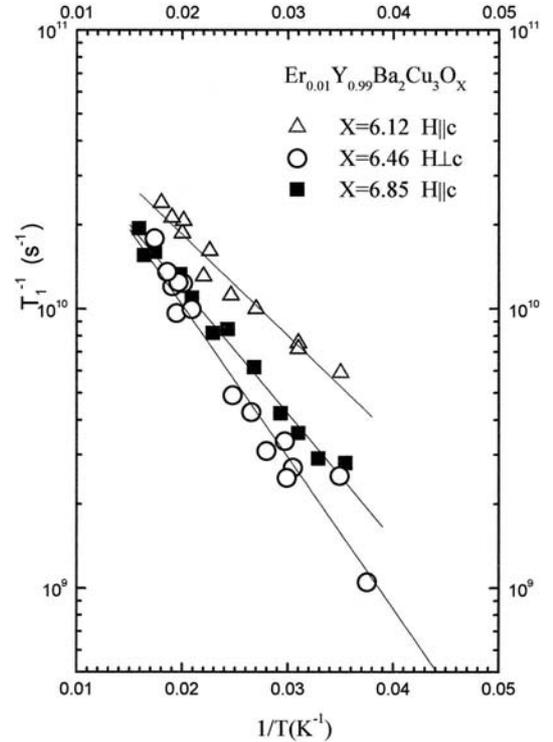

Fig. 3. $Er^{3+}$-SLR rate $T_1^{-1}$ versus $T$ for different oxygen contents $x$. Solid lines are results of fitting procedure by formula (3) using parameters $B_{1,2}$ and $\Delta_{1,2}$ listed in Table 1.

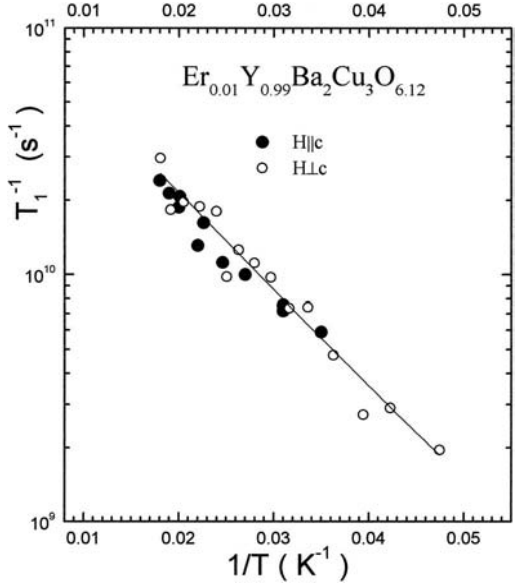

Fig. 4. Temperature dependence of the $Er^{3+}$-SLR rate $T_1^{-1}$ in $Y_{0.99}Er_{0.01}Ba_2Cu_3O_{6.12}$ for $H \parallel c$ (●) and $H \perp c$ (○). Solid line is the best fit by formula (3) with $\Delta_1/k_B = 80$ K and $\Delta_2/k_B = 120$ K.

the Debye energy $k_B \Theta$ ($\Theta \approx 350 \div 400$ K in YBCO) [26] that:

$$T_{1O}^{-1} = \frac{3}{2\pi\rho v^5 \hbar}\left(\frac{\Delta}{\hbar}\right)^3 M^2 \exp(-\Delta/k_B T), \qquad (4)$$

where $v$ is the sound velocity, $\rho$ is the host density, $M$ is the matrix element which couples the ground and excited states of $Er^{3+}$ ion. The product of matrix element $M$ is approximately equal to the splitting of term $^4I_{15/2}$ of $Er^{3+}$ (it was estimated due to INS measurements by Mesot et al. as 80.1 meV [17]) divided to $(N-1)$, where $N=8$ is the number of sublevels of this term. Calculations of $B_{1,2}$ coefficients using formula (4) yield very reasonable values for all of our samples. For example, for $x = 6.12$ using $M \simeq 92.2$ cm$^{-1}$ and the literature data [27] of $\rho = 5.78$ g/cm$^3$, longitudinal and transversal sound velocity $v_l = 4.02 \times 10^5$ cm/s and $v_t = 2.5 \times 10^5$ cm/s, respectively, we can obtain $v \approx 2.813 \times 10^5$ cm/s from the relation

$$\frac{3}{v^3} = \frac{1}{v_l^3} + \frac{2}{v_t^3}, \qquad (5)$$

and further, $B_{1calc} = 3.72 \times 10^{10}$ s$^{-1}$ and $B_{2calc} = 12.2 \times 10^{10}$ s$^{-1}$. Thus, the relationship $(B_{2exp})/(B_{1exp}) \approx (B_{2calc})/(B_{1calc}) \approx ((\Delta_2)/(\Delta_1))^3$, which follows from (4) is also in a good agreement with the data of our measurements and calculations. Hence, our results do not contradict the theory of the electron SLR.

We have neglected in our analysis any contribution from the Korringa-like SLR with a linear dependence of the SLR rate on temperature [28], because its influence on the relaxation processes of $Er^{3+}$ in YBCO is totally masked by the Orbach–Aminov process. Therefore, it is very hard to make any estimations about interactions of host conducting carriers with $Er^{3+}$-ions. The attempt of Shimizu et al. [14–16] to describe the SLR in $Y_{0.97}Er_{0.03}Ba_2Cu_3O_x$ and $La_{1.98-x}Er_{0.02}Sr_xCuO_4$ with the involvement of the Korringa SLR leads to inconsistence with the experimental results on the $T_c$-depression due to Er substitution. It seems to be much useful to study the possible Korringa SLR in cuprate HTS on such doped ions as $Fe^{3+}$, which could be introduced into copper-oxide planes directly [5] or $Gd^{3+}$ [6], suggesting that the corresponding SLR does not reveal a significant contribution of Orbach–Aminov or Raman processes.

In summary, the resonant phonon relaxation processes with the involvement of the lowest CEF excitations of $Er^{3+}$ dominate the electron SLR in $Y_{0.99}Er_{0.01}Ba_2Cu_3O_x$. The experimental data of ESR and SLR measurements are in a good agreement with the values derived from INS studies, related to local phase separation in $ErBa_2Cu_3O_x$. At the same time, only further investigations of the frequency dependence of residual ESR linewidth can clarify peculiarities of magnetic interactions between $Er^{3+}$-ions and $Cu^{2+}$-spin system on copper-oxide planes.

## Acknowledgements

Financial support by the Swiss National Science Foundation under grant No. 7SUPJ048660 is gratefully acknowledged.

## References

[1] M. Mehring, Appl. Magn. Reson. 3 (1992) 383.
[2] N.F. Alekseevski, A.V. Mitin, V.I. Nizhankovskii, I.A. Garifullin, N.N. Garif'yanov, G.G. Khaliullin, E.P. Khlybov, B.I. Kochelaev, L.R. Tagirov, J. Low Temp. Phys. 77 (1989) 87.

[3] B.I. Kochelaev, L. Kan, B. Elschner, E. Elschner, Phys. Rev. B 49 (1994) 13106.
[4] A. Jánossy, J.R. Cooper, L.-C. Brunel, A. Carrington, Phys. Rev. B 50 (1994) 3442.
[5] A.D. Shengelaya, J. Olejniczak, H. Drulis, Physica C 233 (1994) 124.
[6] C. Kessler, M. Mehring, P. Castellaz, G. Borodi, C. Filip, A. Darabont, L.V. Giurgiu, Physica B 229 (1997) 113.
[7] R.Yu. Abdulsabirov, R.Sh. Zhdanov, Ya.S. Isygzon, S.L. Korableva, I.N. Kurkin, L.L. Sedov, I.V. Jasonov, B. Lipold, Supercond. Phys. Chem. Technol. 2 (1989) 52.
[8] V.A. Atsarkin, V.V. Demidov, G.A. Vasneva, Phys. Rev. B 52 (1995) 1290.
[9] V.A. Atsarkin, G.A. Vasneva, V.V. Demidov, JETP 108 (1995) 927.
[10] A.G. Anders, S.V. Volotskii, A.I. Zvyagin, N.M. Chaikovskaya, Fiz. Nizk. Temp. 17 (1991) 637.
[11] M.V. Eremin, I.N. Kurkin, M.P. Rodionova, I.Kh. Salikhov, L.L. Sedov, L.R. Tagirov, Supercond. Phys. Chem. Technol. 4 (1991) 716.
[12] M.X. Huang, J. Barak, S.M. Bhagat, L.C. Gupta, A.K. Rajarajan, R. Vijayaraghavan, J. Appl. Phys. 70 (1991) 5754.
[13] I.N. Kurkin, I.Kh. Salikhov, L.L. Sedov, M.A. Teplov, R.Sh. Zhdanov, JETP 76 (1993) 657.
[14] H. Shimizu, K. Fujiwara, K. Hatada, Physica C 282–287 (1997) 1349.
[15] H. Shimizu, K. Fujiwara, K. Hatada, Physica C 288 (1997) 190.
[16] H. Shimizu, K. Fujiwara, K. Hatada, Physica C 299 (1998) 169.
[17] J. Mesot, P. Allenspach, U. Staub, A. Furrer, Phys. Rev. Lett. 70 (1993) 865.
[18] J. Mesot, P. Allenspach, U. Staub, A. Furrer, H. Mutka, R. Osborn, A. Taylor, Phys. Rev. B 47 (1993) 6027.
[19] U. Staub, J. Mesot, P. Allenspach, A. Furrer, H. Mutka, J. Alloys Comp. 195 (1993) 595.
[20] P. Meufels, B. Rupp, E. Pörschke, Physica C 156 (1988) 441.
[21] A. Abragam, B. Bleaney, EPR of Transition Ions, Clarendon Press, Oxford, 1970.
[22] L. Kan, S. Elschner, B. Elschner, Solid State Commun. 79 (1991) 61.
[23] G.M. Zhidomirov et al., The Interpretation of Complicated EPR Spectra (in Russian), Nauka, Moscow, 1975.
[24] Gh. Cristea, T.L. Bohan, H.J. Stapleton, Phys. Rev. B 4 (1971) 2081.
[25] R. Orbach, H.J. Stapleton, in: S. Geschwind (Ed.), Electron Paramagnetic Resonance, Plenum, New York, 1972.
[26] G.H. Larson, C.D. Jeffries, Phys. Rev. 141 (1966) 461.
[27] H. Ledbetter, J. Mater. Res. 7 (1992) 2905.
[28] J. Korringa, Physica XVI (1950) 601.